\documentclass{jpsj3}
\usepackage{color}
\def\ggs{\buildrel\textstyle > \over {\hbox{\raise0.2ex\hbox{$\sim$}}}}
\def\lls{\buildrel\textstyle < \over {\hbox{\raise0.2ex\hbox{$\sim$}}}}
\def\gsim{\, \lower0.75ex\hbox{$\ggs$}\, }
\def\lsim{\, \lower0.75ex\hbox{$\lls$}\, }
\def \epsilonn{{\epsilon_n}}
\newcommand{\beq}{\begin{equation}}
\newcommand{\eeq}{\end{equation}}

\newcommand{\beqa}{\begin{eqnarray}}
\newcommand{\eeqa}{\end{eqnarray}}
\newcommand{\non}{\nonumber}

\title
{Compact Representation for Specific Heat of Interacting Fermion Systems in Terms of 
Fully Renormalized Matsubara Green Function
}

\author
{Kazumasa Miyake$^{1}$ and Atsushi Tsuruta$^{2}$ 
}

\inst
{
$^{1}$Toyota Physical and Chemical Research Institute, Nagakute, Aichi 480-1192, Japan\\
$^{2}$Division of Materials Physics, Department of Materials Engineering Science,
Graduate School of Engineering Science, Osaka University, Toyonaka, Osaka
560-8531, Japan\\

}

\recdate
{
May 7, 2015
}

\abst
{On the basis of the Luttinger-Ward formalism for the thermodynamic potential, the specific heat 
of single-component interacting fermion systems with fixed chemical potential is compactly expressed 
in terms of the fully renormalized Matsubara Green function. 
}

\begin{document}
\maketitle

\section{Introduction}
It is technically important to obtain a closed-form expression for the specific heat 
of interacting or strongly correlated fermion systems in terms of the 
fully renormalized Matsubara Green function, which is inferred phenomenologically or 
is now feasible on the basis of 
numerical methods in one form or another, such as through dynamical mean-field theory~\cite{DMFT} 
and $1/N$-expansion theory~\cite{Tsuruta}.  In the 1970s, studies on entropy and 
specific heat along this concept were 
published for interacting phonons~\cite{Fulde} and normal Fermi liquids~\cite{Carneiro}. 
Since then, a lot of studies have been published in various fields not only in 
condensed matter physics~\cite{Kita} but also in nuclear physics~\cite{Rios} and 
quark matter physics~\cite{Blaizot}. 
However, to the best of our knowledge, much still needs to be improved.  
For example, 
some contribution is missing in Ref.\ \citen{Fulde} as pointed out in Ref.\ \citen{Carneiro}, 
in which the relevant missing term was calculated from the perturbation in interaction.  

In this paper, for interacting fermion systems, we find that the relevant missing term can be 
compactly expressed in terms of the fully renormalized Matsubara Green function starting from the 
Luttinger-Ward functional for the thermodynamic potential. We show that the specific 
heat can be expressed compactly in terms of the fully renormalized Matsubara Green function.  
In Sect.\ 2, we explicitly derive an expression for the specific heat $C_{\mu}$ 
with fixed chemical potential $\mu$ for a single-component fermion system in the form 
\begin{eqnarray}
& &C_{\mu}=\frac{{\rm d}}{{\rm d}T}\left[
\sum_{\sigma}\frac{1}{N_L}\sum_{\mib k}T\sum_{n}(\varepsilon_{k}-\mu)
G_{{\mib k}\sigma}(i\epsilon_{n})\right]
\nonumber
\\
& &\qquad\qquad\quad
{
+\frac{1}{2}\sum_{\sigma}\frac{1}{N_L}\sum_{\mib{k}}T\sum_{n}\frac{{\rm d}}{{\rm d}T}
\left[\Sigma_{\mib{k}\sigma}(i\epsilon_n)G_{\mib{k}\sigma}(i\epsilon_n)\right],
}
\label{eq:result}
\end{eqnarray}
where $\sigma$ stands for spin degrees of freedom, $N_{L}$ is the number of lattice points, 
$\varepsilon_{k}$ is the dispersion of fermions, and $G_{{\mib k}\sigma}(i\epsilon_{n})$ 
and $\Sigma_{{\mib k}\sigma}(i\epsilon_{n})$ are the 
fully renormalized Matsubara Green function and its self-energy, respectively.  
In Sect.\ 3, some remarks are given.

\section{Derivation of Formula for the Specific Heat}
In this section, we show how the specific heat is calculated 
if the Green function $G(\mib{k},i\epsilon_n)$ of fermions is explicitly given.
We discuss the case of the system whose Hamiltonian is given as 
\begin{eqnarray}
&&H=\sum_{\sigma=\uparrow\downarrow}\sum_{\mib{k}} \varepsilon_{\mib{k}} c^+_{\mib{k}\sigma} 
c_{\mib{k}\sigma}
+
{\frac{1}{2}}
\sum_{\sigma_1\sigma_2}\frac{1}{N_L}\sum_{\mib{k}_1\mib{k}_2\mib{q}}V_{\sigma_1\sigma_2}(\mib{q})c^
+_{\mib{k}_1+\mib{q}\sigma_1} c^
+_{\mib{k}_2-\mib{q}\sigma_2} c_{\mib{k}_2\sigma_2} c_{\mib{k}_1\sigma_1},
\label{eq:A1}
\end{eqnarray}
where $N_L$ is the number of lattice points.  
According to Luttinger and Ward\cite{Luttinger,Abrikosov}, the thermodynamic potential 
$\Omega$ is given as
\begin{eqnarray}
\Omega=\Omega_0+\tilde{\Omega}+\Omega',
\label{eq:A2}
\end{eqnarray}
where
\begin{eqnarray}
\Omega_0&=&\sum_\sigma T\sum_{n}\frac{1}{N_L}\sum_{\mib{k}}e^{i\epsilon_n\delta}
\ln\left[G^{0}_{{\mib{k}}\sigma}(i\epsilon_n)\right]\label{A3},
\label{eq:A3}
\end{eqnarray}
and
\begin{eqnarray}
\tilde{\Omega}&=&-\sum_\sigma T\sum_{n}\frac{1}{N_L}\sum_{\mib{k}}
\left\{\ln\left[1-G^{0}_{{\mib{k}}\sigma}(i\epsilon_n)\Sigma_{{\mib{k}}\sigma}(i\epsilon_n)\right]
+\Sigma_{{\mib{k}}\sigma}(i\epsilon_n)G_{{\mib{k}}\sigma}(i\epsilon_n)\right\},
\label{eq:A4}
\end{eqnarray}
where $G^{0}_{{\mib k}\sigma}(i\epsilonn)$ is the unperturbed Green function
\begin{equation}
G^{0}_{{\mib k}\sigma}(i\epsilonn)=\frac{1}{i\epsilonn-\varepsilon_{\mib{k}}+\mu},
\label{eq:A5}
\end{equation}
and $\Omega'$ is the perturbation expansion for the thermodynamic potential in terms of 
the skeleton diagrams with the fully renormalized Green function $G$ and the bare 
two-body interaction $V$.  
The explicit form of Feynman diagrams for $\Omega'$ is shown in Fig. 1, and its analytic form is 
given by 
\begin{eqnarray}
\Omega'&=&
{
-\frac{1}{2}\sum_{\sigma,\sigma^{\prime}} T\sum_{n_1}T\sum_{n_2}
\frac1{N_L^2}\sum_{\mib{k}_1\mib{k}_2}
G_{\mib{k}_{1}\sigma}(i\epsilon_{n_1})e^{i\epsilon_{n_{1}}\delta}
G_{\mib{k}_{2}\sigma}(i\epsilon_{n_2})e^{i\epsilon_{n_{2}}\delta}
V_{\sigma\sigma^{\prime}}(0)
}
\non
\\
&&
+
{\frac{1}{2}}\sum_\sigma T\sum_{n_1}T\sum_{n_2}
\frac1{N_L^2}\sum_{\mib{k}_1\mib{k}_2}
G_{\mib{k}_{1}\sigma}(i\epsilon_{n_1})
G_{\mib{k}_{2}\sigma}(i\epsilon_{n_2})
V_{\sigma\sigma}(\mib{k}_1-\mib{k}_2)\non
\\
&&-
{\frac{1}{4}}
\sum_\sigma T\sum_{n_1}T\sum_{n_2}T\sum_{n_3}
\frac1{N_L^3}\sum_{\mib{k}_1\sim\mib{k}_3}
G_{\mib{k}_{1}\sigma}(i\epsilon_{n_1})
G_{\mib{k}_{2}\sigma}(i\epsilon_{n_2})
G_{\mib{k}_{3}\sigma}(i\epsilon_{n_3})\non
\\
&&\qquad\times G_{\mib{k}_1-\mib{k}_2+\mib{k}_{3},\sigma}(i\epsilon_{n_1}-i\epsilon_{n_2}+i\epsilon_{n_3})
V_{\sigma\sigma}(\mib{k}_1-\mib{k}_2)
V_{\sigma\sigma}(\mib{k}_2-\mib{k}_3)
\non
\\
&&+
{\frac{1}{4}}
\sum_{\sigma,\sigma^{\prime}} T\sum_{n_1}T\sum_{n_2}T\sum_{n_3}
\frac1{N_L^3}\sum_{\mib{k}_1\sim\mib{k}_3}
G_{\mib{k}_{1}\sigma}(i\epsilon_{n_1})
G_{\mib{k}_{2}\sigma}(i\epsilon_{n_2})
G_{\mib{k}_{3}\sigma^{\prime}}(i\epsilon_{n_3})\non
\\
&&\qquad\times G_{\mib{k}_1-\mib{k}_2+\mib{k}_{3},\sigma^{\prime}}
(i\epsilon_{n_1}-i\epsilon_{n_2}+i\epsilon_{n_3})
[V_{\sigma\sigma^{\prime}}(\mib{k}_1-\mib{k}_2)]^2\non\\
&&+
{\frac{1}{6}}
\sum_\sigma T\sum_{n_1}T\sum_{n_2}T\sum_{n_3}T\sum_{n_4}
\frac1{N_L^4}\sum_{\mib{k}_1\sim\mib{k}_4}
G_{\mib{k}_{1}\sigma}(i\epsilon_{n_1})
G_{\mib{k}_{2}\sigma}(i\epsilon_{n_2})
G_{\mib{k}_{3}\sigma}(i\epsilon_{n_3})
G_{\mib{k}_{4}\sigma}(i\epsilon_{n_4})\non
\\
&&\qquad\times G_{\mib{k}_1-\mib{k}_2+\mib{k}_{4},\sigma}
(i\epsilon_{n_1}-i\epsilon_{n_2}+i\epsilon_{n_4})
{G_{\mib{k}_1-\mib{k}_3+\mib{k}_{4},\sigma}(i\epsilon_{n_1}-i\epsilon_{n_3}+i\epsilon_{n_4})}
\non
\\
&&\qquad{\times 
V_{\sigma\sigma}(\mib{k}_1-\mib{k}_2)
V_{\sigma\sigma}(\mib{k}_2-\mib{k}_3)
V_{\sigma\sigma}(\mib{k}_3-\mib{k}_4)+\cdots\cdots},
\label{eq:A6}
\end{eqnarray}
where the part indicated by dots includes the remaining terms of the third-order perturbation 
and all the higher-order perturbation terms.  
{We should note here that the factor 2 in the denominator of each perturbation 
term of $\Omega'$ is not explicitly mentioned in the textbook of Ref.\ \citen{Abrikosov}.}
Therefore, $\Omega'/T$ is given as
\begin{eqnarray}
\frac{\Omega'}{T}&=&
{
-\frac{1}{2}\sum_{\sigma,\sigma^{\prime}}T\sum_{n_1}\sum_{n_2}
\frac1{N_L^2}\sum_{\mib{k}_1\mib{k}_2}
G_{\mib{k}_{1}\sigma}(i\epsilon_{n_1})e^{i\epsilon_{n_{1}}\delta}
G_{\mib{k}_{2}\sigma}(i\epsilon_{n_2})e^{i\epsilon_{n_{2}}\delta}
V_{\sigma\sigma^{\prime}}(0)}
\non
\\
&&
+{\frac{1}{2}}
\sum_\sigma T\sum_{n_1}\sum_{n_2}
\frac1{N_L^2}\sum_{\mib{k}_1\mib{k}_2}
G_{\mib{k}_{1}\sigma}(i\epsilon_{n_1})
G_{\mib{k}_{2}\sigma}(i\epsilon_{n_2})
V_{\sigma\sigma}(\mib{k}_1-\mib{k}_2)\non\\
&&+
{\frac{1}{4}}
\sum_{\sigma,\sigma^{\prime}} T^2\sum_{n_1}\sum_{n_2}\sum_{n_3}
\frac1{N_L^3}\sum_{\mib{k}_1\sim\mib{k}_3}
G_{\mib{k}_{1}\sigma}(i\epsilon_{n_1})
G_{\mib{k}_{2}\sigma}(i\epsilon_{n_2})
G_{\mib{k}_{3}\sigma^{\prime}}(i\epsilon_{n_3})\non\\
&&\qquad\times G_{\mib{k}_1-\mib{k}_2+\mib{k}_{3},\sigma^{\prime}}
(i\epsilon_{n_1}-i\epsilon_{n_2}+i\epsilon_{n_3})\non\\
&&\qquad\times\left\{
[V_{\sigma{\sigma}^{\prime}}(\mib{k}_1-\mib{k}_2)]^2
-\delta_{\sigma\sigma^{\prime}}V_{\sigma\sigma^{\prime}}(\mib{k}_1-\mib{k}_2)
V_{\sigma\sigma^{\prime}}(\mib{k}_2-\mib{k}_3)
\right\}\non\\
&&+
{\frac{1}{6}}
\sum_\sigma T^3\sum_{n_1}\sum_{n_2}\sum_{n_3}\sum_{n_4}
\frac1{N_L^4}\sum_{\mib{k}_1\sim\mib{k}_4}
G_{\mib{k}_{1}\sigma}(i\epsilon_{n_1})
G_{\mib{k}_{2}\sigma}(i\epsilon_{n_2})
G_{\mib{k}_{3}\sigma}(i\epsilon_{n_3})
G_{\mib{k}_{4}\sigma}(i\epsilon_{n_4})\non\\
&&\qquad\times G_{\mib{k}_1-\mib{k}_2+\mib{k}_{4},\sigma}
(i\epsilon_{n_1}-i\epsilon_{n_2}+i\epsilon_{n_4})
{G_{\mib{k}_1-\mib{k}_3+\mib{k}_{4},\sigma}(i\epsilon_{n_1}-i\epsilon_{n_3}+i\epsilon_{n_4})}
\non\\
&&\qquad{\times V_{\sigma\sigma}(\mib{k}_1-\mib{k}_2)
V_{\sigma\sigma}(\mib{k}_2-\mib{k}_3)
V_{\sigma\sigma}(\mib{k}_3-\mib{k}_4)+\cdots\cdots}.
\label{eq:A7}
\end{eqnarray}

\begin{figure}
\includegraphics[width=12cm]{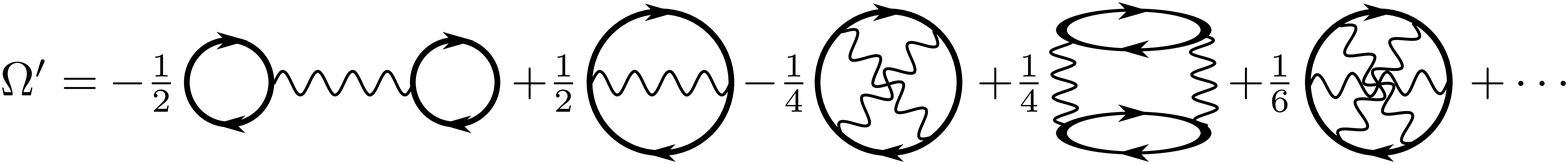}
\caption{Feynman diagram representation of the thermodynamic potential $\Omega'$. 
The part indicated by dots includes the remaining terms of the third-order perturbation and all the 
higher-order perturbation terms.}
\end{figure}

First, we differentiate this term with respect to an explicit temperature dependence other than 
that included in the fully renormalized Green function $G_{\mib{k}_{j}}(i\epsilon_{n_{j}})$:
\begin{eqnarray}
\frac{\partial}{\partial T}\left(\frac{\Omega'}{T}\right)&=&
{
-\frac{1}{2}\sum_{\sigma,\sigma^{\prime}}\sum_{n_1}\sum_{n_2}
\frac1{N_L^2}\sum_{\mib{k}_1\mib{k}_2}
G_{\mib{k}_{1}\sigma}(i\epsilon_{n_1})e^{i\epsilon_{n_{1}}\delta}
G_{\mib{k}_{2}\sigma}(i\epsilon_{n_2})e^{i\epsilon_{n_{2}}\delta}
V_{\sigma\sigma^{\prime}}(0)}
\non
\\
&&
+{\frac{1}{2}}
\sum_\sigma \sum_{n_1}\sum_{n_2}
\frac1{N_L^2}\sum_{\mib{k}_1\mib{k}_2}
G_{\mib{k}_{1}\sigma}(i\epsilon_{n_1})
G_{\mib{k}_{2}\sigma}(i\epsilon_{n_2})
V_{\sigma\sigma}(\mib{k}_1-\mib{k}_2)\non\\
&&+{\frac{1}{2}}
\sum_{\sigma,\sigma^{\prime}} T\sum_{n_1}\sum_{n_2}\sum_{n_3}
\frac1{N_L^3}\sum_{\mib{k}_1\sim\mib{k}_3}
G_{\mib{k}_{1}\sigma}(i\epsilon_{n_1})
G_{\mib{k}_{2}\sigma}(i\epsilon_{n_2})
G_{\mib{k}_{3},\sigma^{\prime}}(i\epsilon_{n_3})\non\\
&&\qquad\times G_{\mib{k}_1-\mib{k}_2+\mib{k}_{3},\sigma^{\prime}}
(i\epsilon_{n_1}-i\epsilon_{n_2}+i\epsilon_{n_3})\non\\
&&\qquad\times\left\{
[V_{\sigma{\sigma}^{\prime}}(\mib{k}_1-\mib{k}_2)]^2
-\delta_{\sigma\sigma^{\prime}}V_{\sigma\sigma^{\prime}}(\mib{k}_1-\mib{k}_2)
V_{\sigma\sigma^{\prime}}(\mib{k}_2-\mib{k}_3)
\right\}\non\\
&&+{\frac{1}{2}}
\sum_\sigma T^2\sum_{n_1}\sum_{n_2}\sum_{n_3}\sum_{n_4}
\frac1{N_L^4}\sum_{\mib{k}_1\sim\mib{k}_4}
G_{\mib{k}_{1}\sigma}(i\epsilon_{n_1})
G_{\mib{k}_{2}\sigma}(i\epsilon_{n_2})
G_{\mib{k}_{3}\sigma}(i\epsilon_{n_3})
G_{\mib{k}_{4}\sigma}(i\epsilon_{n_4})\non\\
&&\qquad\times G_{\mib{k}_1-\mib{k}_2+\mib{k}_{4},\sigma}(i\epsilon_{n_1}-i\epsilon_{n_2}+i\epsilon_{n_4})
{G_{\mib{k}_1-\mib{k}_3+\mib{k}_{4},\sigma}(i\epsilon_{n_1}-i\epsilon_{n_3}+i\epsilon_{n_4})}
\non\\
&&\qquad{\times V_{\sigma\sigma}(\mib{k}_1-\mib{k}_2)
V_{\sigma\sigma}(\mib{k}_2-\mib{k}_3)
V_{\sigma\sigma}(\mib{k}_3-\mib{k}_4)+\cdots\cdots}.
\label{eq:A8}
\\
&=&
{\frac{1}{2T^2}}
\sum_{\sigma}T\sum_n\sum_{\mib{k}}\Sigma_{\mib{k}\sigma}(i\epsilon_n)
G_{\mib{k}\sigma}(i\epsilon_n).
\label{eq:A9}
\end{eqnarray}
Note that all the differentiations in the present paper are performed with the chemical 
potential $\mu$ fixed.  
In deriving Eq.\ (\ref{eq:A9}) from Eq.\ (\ref{eq:A8}), we have used an explicit form of 
the perturbation expansion of the self-energy, as discussed in Refs.\ \citen{Luttinger} 
and \citen{Abrikosov}.
On the other hand, with the use of the identity
\begin{eqnarray}
\frac{\partial}{\partial T}\left(\frac{\Omega'}{T}\right)&=&
\frac{1}{T}\frac{\partial \Omega'}{\partial T}-\frac{1}{T^2}\Omega'
\end{eqnarray}
and Eq.\ (\ref{eq:A9}), we arrive at the following relation:
\begin{eqnarray}
\frac{\partial \Omega'}{\partial T}&=&\frac{\Omega'}{T}+{\frac{1}{2T}}
\sum_{\sigma}T\sum_n\frac{1}{N_L}\sum_{\mib{k}}\Sigma_{\mib{k}\sigma}(i\epsilon_n)
G_{\mib{k}\sigma}(i\epsilon_n).
\label{eq:A11}
\end{eqnarray}
%Here, we remark that this explicit $T$ dependence of $\Omega'$ has been overlooked in 
%Refs.\ \citen{Fulde} and \citen{Kita}. 

Since the second term of $\Omega$, in Eq.\ (\ref{eq:A2}), is given by a compact form in terms of 
the fully renormalized Green function $G$ and the self-energy $\Sigma$, the 
remaining problem in calculating the 
specific heat in terms of $G$ and $\Sigma$ is how to extract information included in $\Omega'$.
According to the theory of Luttinger and Ward\cite{Luttinger,Abrikosov}, 
the variation of $\Omega'$ with respect to the fully renormalized Green function satisfies the 
following variational relation:
\begin{eqnarray}
\delta\Omega'=\sum_\sigma T\sum_n\frac1{N_L}\sum_{\mib{k}}
\Sigma_{\mib{k}\sigma}(i\epsilon_n)\delta G_{\mib{k}\sigma}(i\epsilon_n).
\label{eq:A12}
\end{eqnarray}
Therefore, the entropy $S'$ arising from $\Omega'$ is expressed as
\begin{eqnarray}
S'&=&
-\frac{{\rm d}\Omega'}{{\rm d}T}\non\\
&=&-\sum_\sigma T\sum_{n}\frac{1}{N_L}\sum_{\mib{k}}
\Sigma_{\mib{k}\sigma}(i\epsilon_n)\frac{{\rm d} G_{\mib{k}\sigma}(i\epsilon_n)}{{\rm d}T}
-\frac{\partial \Omega'}{\partial T}, 
\label{eq:A12a}
\end{eqnarray}
where ${\rm d}/{\rm d}T$ implies the total differentiation through the explicit $T$ dependence of 
$\Omega'$ and through those of $G$ and $\Sigma$,
while $\partial/\partial T$ implies a differentiation only through the explicit $T$ dependence.
With the use of Eq.\ (\ref{eq:A11}), Eq.\ (\ref{eq:A12a}) is reduced to 
\begin{equation}
S'=
-\sum_\sigma T\sum_{n}\frac{1}{N_L}\sum_{\mib{k}}
\Sigma_{\mib{k}\sigma}(i\epsilon_n)\frac{{\rm d}G_{\mib{k}\sigma}(i\epsilon_n)}{{\rm d}T}
-\frac{\Omega'}{T}
-{\frac{1}{2}}
\sum_{\sigma}\sum_n
{\frac{1}{N_{L}}}
\sum_{\mib{k}}\Sigma_{\mib{k}\sigma}(i\epsilon_n)
G_{\mib{k}\sigma}(i\epsilon_n).
\label{eq:A13}
\end{equation}
Hereafter, for concise presentation, we abbreviate the spin variables $\sigma$ in $G$ and $\Sigma$, 
considering the case of a paramagnetic state. 

Then, the specific heat $C'$ arising from $\Omega'$ is given as
\begin{eqnarray}
C'&=&T\frac{{\rm d}S'}{{\rm d}T}\non
\\
&=&-\sum_\sigma T\left[
T\sum_n\frac{1}{N_L}\sum_{\mib{k}}\frac{{\rm d}}{{\rm d}T}
\left(\Sigma_{\mib{k}}(i\epsilon_n)\frac{{\rm d} G_{\mib{k}}(i\epsilon_n)}{{\rm d}T}\right)
+\sum_n\frac{1}{N_L}\sum_{\mib{k}}\Sigma_{\mib{k}}(i\epsilon_n)
\frac{{\rm d} G_{\mib{k}}(i\epsilon_n)}{{\rm d}T}
\right]\non
\\
&&\qquad -T\left\{
\frac1T{\left(\frac{{\rm d}\Omega'}{{\rm d}T}\right)_{*}}
+\frac{\partial}
{\partial T}\left(\frac{\Omega'}{T}\right)
+{\frac{1}{2}}
\sum_{n}\frac{1}{N_L}\sum_{\mib{k}}\frac{d}{{\rm d}T}
\left[\Sigma_{\mib{k}}(i\epsilon_n)G_{\mib{k}}(i\epsilon_n)\right]
\right\}
\non
\\
&=&-\sum_\sigma T\sum_n\frac1{N_L}\sum_{\mib{k}}
\left\{
T\frac{d}{{\rm d}T}\left[\Sigma_{\mib{k}}(i\epsilon_n)
\frac{{\rm d} G_{\mib{k}}(i\epsilon_n)}{{\rm d}T}\right]
+2\Sigma_{\mib{k}}(i\epsilon_n)\frac{{\rm d} G_{\mib{k}}(i\epsilon_n)}{{\rm d}T}\right.\non
\\
&&\qquad\qquad\qquad\qquad\qquad\quad
\left.+\frac1T\Sigma_{\mib{k}}(i\epsilon_n)G_{\mib{k}}(i\epsilon_n)
+\frac{d}{{\rm d}T}
\left[\Sigma_{\mib{k}}(i\epsilon_n)G_{\mib{k}}(i\epsilon_n)\right]
\right\}
\non
\\
&&\qquad
{+\frac{1}{2}}{\sum_{\sigma}}
{T\sum_{n}\frac{1}{N_L}\sum_{\mib{k}}\frac{d}{{\rm d}T}
\left[\Sigma_{\mib{k}}(i\epsilon_n)G_{\mib{k}}(i\epsilon_n)\right]}
.\label{c1}
\end{eqnarray}
{Note that $({\rm d}\Omega^{\prime}/{\rm d}T)_{*}$ in the second equality of Eq.\ (\ref{c1}) 
is the $T$-derivative of $\Omega^{\prime}$ only through the $T$ dependence of 
$G_{\mib{k}}(i\epsilon_n)$ 
while $\partial(\Omega^{\prime}/T)/\partial T$ is that through the explicit $T$ dependence of 
$(\Omega^{\prime}/T)$ as in Eqs.\ (\ref{eq:A8}) and (\ref{eq:A9}). }
In deriving the last equality of Eq.\ (\ref{c1}), we have used 
Eqs.\ (\ref{eq:A9}) and (\ref{eq:A12}).  
{For the discussion below, we define the first and second terms of Eq.\ (\ref{c1}) as 
$C^{\prime}_{1}$ and $C^{\prime}_{2}$, respectively:
\begin{eqnarray}
& &C^{\prime}_{1}\equiv 
-\sum_\sigma T\sum_n\frac1{N_L}\sum_{\mib{k}}
\left\{
T\frac{d}{{\rm d}T}\left[\Sigma_{\mib{k}}(i\epsilon_n)
\frac{{\rm d} G_{\mib{k}}(i\epsilon_n)}{{\rm d}T}\right]
+2\Sigma_{\mib{k}}(i\epsilon_n)\frac{{\rm d} G_{\mib{k}}(i\epsilon_n)}{{\rm d}T}\right.\non
\\
&&\qquad\qquad\qquad\qquad\qquad\qquad
\left.+\frac1T\Sigma_{\mib{k}}(i\epsilon_n)G_{\mib{k}}(i\epsilon_n)
+\frac{d}{{\rm d}T}
\left[\Sigma_{\mib{k}}(i\epsilon_n)G_{\mib{k}}(i\epsilon_n)\right]
\right\},
\label{c11}
\end{eqnarray}
and 
\begin{equation}
C^{\prime}_{2}\equiv
\frac{1}{2}\sum_{\sigma}\frac{1}{N_L}\sum_{\mib{k}}T\sum_{n}\frac{{\rm d}}{{\rm d}T}
\left[\Sigma_{\mib{k}\sigma}(i\epsilon_n)G_{\mib{k}\sigma}(i\epsilon_n)\right].
\label{c12}
\end{equation}
}

The contribution to the specific heat from the thermodynamic potential ${\tilde \Omega}$, 
given by Eq.\ (\ref{eq:A4}), is calculated as follows. 
With the use of the definition in Eq.\ (\ref{eq:A4}) and the relation 
$1-G^0_{\mib{k}}(i\epsilonn)\Sigma_{\mib{k}}(i\epsilonn)=G^0_{\mib{k}}(i\epsilonn)
G_{\mib{k}}^{-1}(i\epsilonn),$ the entropy $\tilde{S}$ from $\tilde{\Omega}$ is expressed as
\begin{eqnarray}
\tilde{S}=-\frac{{\rm d} \tilde{\Omega}}{{\rm d} T}
&=&\sum_\sigma\sum_{n}\frac{1}{N_L}\sum_{\mib{k}}
\left\{\ln\left[ G^0_{\mib{k}}(i\epsilon_n)G^{-1}_{\mib{k}}(i\epsilon_n)\right]
+\Sigma_{\mib{k}}(i\epsilonn)G_{\mib{k}}(i\epsilonn)\right\}\non
\\
&&+\sum_\sigma T\sum_{n}\frac{1}{N_L}\sum_{\mib{k}}\left[
\frac{1}{G^0_{\mib{k}}(i\epsilon_n)}\frac{{\rm d} G^0_{\mib{k}}(i\epsilon_n)}{{\rm d}T}
+G_{{\mib{k}}}
(i\epsilon_n)\frac{{\rm d} G^{-1}_{\mib{k}}(i\epsilon_n)}{{\rm d}T}\right.\non
\\
&&\qquad\qquad\qquad\qquad\qquad
\left.+\frac{d\Sigma_{\mib{k}}(i\epsilon_n)}{{\rm d}T}G_{\mib{k}}(i\epsilon_n)
+\Sigma_{\mib{k}}(i\epsilon_n)\frac{{\rm d}G _{\mib{k}}(i\epsilon_n)}{{\rm d}T}\right].
\end{eqnarray}
Therefore, the specific heat $\tilde{C}$ from $\tilde{\Omega}$ is expressed as
\begin{eqnarray}
&&\tilde{C}=T\frac{d\tilde{S}}{{\rm d}T}\non\\
&&
\quad
=2\sum_\sigma T\sum_{n}\frac{1}{N_L}\sum_{\mib{k}}\left[
\frac{1}{G^0_{\mib{k}}(i\epsilon_n)}\frac{{\rm d} G^0_{\mib{k}}(i\epsilon_n)}{{\rm d}T}
+G_{\mib{k}}(i\epsilon_n)
\frac{{\rm d} G^{-1}_{\mib{k}}(i\epsilon_n)}{{\rm d}T}\right.\non\\
&&
\qquad\qquad\qquad\qquad\qquad\quad
\left.+\frac{d\Sigma_{\mib{k}}(i\epsilon_n)}{{\rm d}T}G_{\mib{k}}(i\epsilon_n)
+\Sigma_{\mib{k}}(i\epsilon_n)\frac{{\rm d}G _{\mib{k}}(i\epsilon_n)}{{\rm d}T}\right]\non\\
&&
\quad\quad
+\sum_\sigma T^2\sum_{n}\frac{1}{N_L}\sum_{\mib{k}}\frac{d}{{\rm d}T}\left[
\frac{1}{G^0(i\epsilon_n)}\frac{{\rm d} G^0_{\mib{k}}(i\epsilon_n)}{{\rm d}T}
+G_{\mib{k}}(i\epsilon_n)
\frac{{\rm d} G^{-1}_{\mib{k}}(i\epsilon_n)}{{\rm d}T}\right.\non\\
&&
\qquad\qquad\qquad\qquad\qquad\qquad
\left.+\frac{d\Sigma_{\mib{k}}(i\epsilon_n)}{{\rm d}T}G_{\mib{k}}(i\epsilon_n)
+\Sigma_{\mib{k}}(i\epsilon_n)\frac{{\rm d}G _{\mib{k}}(i\epsilon_n)}{{\rm d}T}\right].
\label{c2}
\end{eqnarray}

%%%%%%%% '±'±'É 2ch_06_Suppl.tex 'ð'}"ü %%%%%%%%%
According to the definition of the self-energy $\Sigma$, i.e., $G^{-1}\equiv(G^{0})^{-1}-\Sigma$, and 
$(G^{0})^{-1}\equiv i\epsilon_{n}-\epsilon_{k}+\mu$, the following relations hold:
\begin{equation}
\frac{{\rm d}G_{\mib k}^{-1}(i\epsilon_{n})}{{\rm d}T}=\frac{i\epsilon_{n}}{T}-
\frac{{\rm d}\Sigma_{\mib k}(i\epsilon_{n})}{{\rm d}T},
\label{eq:S1}
\end{equation}
and 
\begin{equation}
\frac{{\rm d}G^{0}_{\mib k}(i\epsilon_{n})}{{\rm d}T}=-\frac{i\epsilon_{n}}{T}
\left[G^{0}_{\mib k}(i\epsilon_{n})\right]^{2}.
\label{eq:S2}
\end{equation}
With the use of these relations, Eq.\ (\ref{c2}) is reduced to 
\begin{eqnarray}
& &
{\tilde C}=\sum_{\sigma}T\sum_{n}\frac{1}{N_L}\sum_{\mib k}
\left[-\frac{2i\epsilon_{n}}{T}G^{0}_{\mib k}(i\epsilon_{n})
+\frac{2i\epsilon_{n}}{T}G_{\mib k}(i\epsilon_{n})
+2\Sigma_{\mib k}(i\epsilon_{n})\frac{{\rm d}G_{\mib k}(i\epsilon_{n})}{{\rm d}T}
\right]
\nonumber
\\
& &
\qquad
+\sum_{\sigma}T\sum_{n}\frac{1}{N_L}\sum_{\mib k}
T\frac{{\rm d}}{{\rm d}T}
\left[-\frac{i\epsilon_{n}}{T}G^{0}_{\mib k}(i\epsilon_{n})
+\frac{i\epsilon_{n}}{T}G_{\mib k}(i\epsilon_{n})
+\Sigma_{\mib k}(i\epsilon_{n})\frac{{\rm d}G_{\mib k}(i\epsilon_{n})}{{\rm d}T}
\right].
\label{eq:S3}
\end{eqnarray}
The first term in the brace of Eq.\ (\ref{c11}) and the last term in the second bracket of 
Eq.\ (\ref{eq:S3}) are of the same size and have opposite signs, 
so that they cancel each other.  
The second term in the brace of Eq.\ (\ref{c11}) and the last term in the first bracket of 
Eq.\ (\ref{eq:S3}) 
are also of the same size and have opposite signs, so that they also cancel each other.  
Therefore, by adding Eqs.\ (\ref{c1}) and (\ref{eq:S3}), 
${C^{\prime}_{1}}+{\tilde C}$ is expressed as 
\begin{eqnarray}
& &
{C^{\prime}_{1}}+{\tilde C}=\sum_{\sigma}\sum_{n}\frac{1}{N_L}\sum_{\mib k}
\left\{
-\frac{{\rm d}}{{\rm d}T}\left[TG_{\mib k}(i\epsilon_{n})\Sigma_{\mib k}(i\epsilon_{n})\right]
+2i\epsilon_{n}\left[G_{\mib k}(i\epsilon_{n})-G^{0}_{\mib k}(i\epsilon_{n})\right]
\right.
\nonumber
\\
& &
\left.
\qquad\qquad\qquad\qquad\qquad\qquad
+T^{2}\frac{{\rm d}}{{\rm d}T}\left[\frac{i\epsilon_{n}}{T}
\left(G_{\mib k}(i\epsilon_{n})-G^{0}_{\mib k}(i\epsilon_{n})\right)
\right]
\right\}.
\label{eq:S4}
\end{eqnarray}
Because of the identity $G-G^{0}=G^{0}G\Sigma$ and the relation 
${\rm d}(i\epsilon_{n}/T)/{\rm d}T=0$, Eq.\ (\ref{eq:S4}) is reduced to 
\begin{eqnarray}
& &
{C^{\prime}_{1}}+{\tilde C}=\sum_{\sigma}\sum_{n}\frac{1}{N_L}\sum_{\mib k}
\left\{
-\frac{{\rm d}}{{\rm d}T}\left[TG_{\mib k}(i\epsilon_{n})\Sigma_{\mib k}(i\epsilon_{n})\right]
+2i\epsilon_{n}G^{0}_{\mib k}(i\epsilon_{n})G_{\mib k}(i\epsilon_{n})\Sigma_{\mib k}(i\epsilon_{n})
\right.
\nonumber
\\
& &
\left.
\qquad\qquad\qquad\qquad\qquad\qquad
+T\,i\epsilon_{n}\frac{{\rm d}}{{\rm d}T}\left[
G^{0}_{\mib k}(i\epsilon_{n})G_{\mib k}(i\epsilon_{n})\Sigma_{\mib k}(i\epsilon_{n})
\right]
\right\}.
\label{eq:S5}
\end{eqnarray}
With the use of Eq.\ (\ref{eq:S2}), the last term in the brace of Eq.\ (\ref{eq:S5}) 
is transformed to 
\begin{eqnarray}
& &
T\, i\epsilon_{n}\frac{{\rm d}}{{\rm d}T}\left[G^{0}_{\mib k}(i\epsilon_{n})G_{\mib k}(i\epsilon_{n})
\Sigma_{\mib k}(i\epsilon_{n})\right]
\nonumber
\\
& &
\qquad
=-(i\epsilon_{n})^{2}\left[G^{0}_{\mib k}(i\epsilon_{n})\right]^{2}
G_{\mib k}(i\epsilon_{n})\Sigma_{\mib k}(i\epsilon_{n})
+T\, i\epsilon_{n}G^{0}_{\mib k}(i\epsilon_{n})\frac{{\rm d}}{{\rm d}T}
\left[G_{\mib k}(i\epsilon_{n})\Sigma_{\mib k}(i\epsilon_{n})\right].
\label{eq:S6}
\end{eqnarray}
Therefore, Eq.\ (\ref{eq:S5}) is reduced to 
\begin{eqnarray}
& &
{C^{\prime}_{1}}+{\tilde C}=\sum_{\sigma}\sum_{n}\frac{1}{N_L}\sum_{\mib k}
\left\{
-\frac{{\rm d}}{{\rm d}T}\left[TG_{\mib k}(i\epsilon_{n})\Sigma_{\mib k}(i\epsilon_{n})\right]
+i\epsilon_{n}G^{0}_{\mib k}(i\epsilon_{n})G_{\mib k}(i\epsilon_{n})\Sigma_{\mib k}(i\epsilon_{n})
\right.
\nonumber
\\
& &
\left.
\qquad\qquad\qquad\qquad\qquad\qquad
+i\epsilon_{n}TG^{0}_{\mib k}(i\epsilon_{n})\frac{{\rm d}}{{\rm d}T}
\left[G_{\mib k}(i\epsilon_{n})\Sigma_{\mib k}(i\epsilon_{n})\right]
\right.
\nonumber
\\
& &
\left.
\qquad\qquad\qquad\qquad\qquad\qquad
+i\epsilon_{n}G^{0}_{\mib k}(i\epsilon_{n})
\left[1-i\epsilon_{n}G^{0}_{\mib k}(i\epsilon_{n})\right]
G_{\mib k}(i\epsilon_{n})\Sigma_{\mib k}(i\epsilon_{n})
\right\}.
\label{eq:S7}
\end{eqnarray}
With the use of the relations $i\epsilon_{n}G^{0}_{\mib k}(i\epsilon_{n})
=1+(\varepsilon_{k}-\mu)G^{0}_{\mib k}(i\epsilon_{n})$ and 
$i\epsilon_{n}[G^{0}_{\mib k}(i\epsilon_{n})]^{2}=-T[{\rm d}G^{0}_{\mib k}(i\epsilon_{n})/{\rm d}T]$, 
which is equivalent to Eq.\ (\ref{eq:S2}), the right-hand side of Eq.\ (\ref{eq:S7}) 
is rearranged as  
\begin{eqnarray}
& &
{C^{\prime}_{1}}+{\tilde C}=\sum_{\sigma}\sum_{n}\frac{1}{N_L}\sum_{\mib k}
\left\{
-\frac{{\rm d}}{{\rm d}T}\left[TG_{\mib k}(i\epsilon_{n})\Sigma_{\mib k}(i\epsilon_{n})\right]
+i\epsilon_{n}G^{0}_{\mib k}(i\epsilon_{n})\frac{{\rm d}}{{\rm d}T}
\left[TG_{\mib k}(i\epsilon_{n})\Sigma_{\mib k}(i\epsilon_{n})\right]
\right.
\nonumber
\\
& &
\left.
\qquad\qquad\qquad\qquad\qquad\qquad
-i\epsilon_{n}(\varepsilon_{k}-\mu)\left[G^{0}_{\mib k}(i\epsilon_{n})\right]^{2}
G_{\mib k}(i\epsilon_{n})\Sigma_{\mib k}(i\epsilon_{n})
\right\}
\nonumber
\\
& &
\qquad\quad
=\sum_{\sigma}\sum_{n}\frac{1}{N_L}\sum_{\mib k}
\left\{
-\frac{{\rm d}}{{\rm d}T}\left[TG_{\mib k}(i\epsilon_{n})\Sigma_{\mib k}(i\epsilon_{n})\right]
\right.
\nonumber
\\
& &
\left.
\qquad\qquad\qquad\qquad\qquad\qquad
+\left[1+(\varepsilon_{k}-\mu)G^{0}_{\mib k}(i\epsilon_{n})\right]
\frac{{\rm d}}{{\rm d}T}\left[TG_{\mib k}(i\epsilon_{n})\Sigma_{\mib k}(i\epsilon_{n})\right]
\right.
\nonumber
\\
& &
\left.
\qquad\qquad\qquad\qquad\qquad\qquad
+T(\varepsilon_{k}-\mu)\frac{{\rm d}G^{0}_{\mib k}(i\epsilon_{n})}{{\rm d}T}
G_{\mib k}(i\epsilon_{n})\Sigma_{\mib k}(i\epsilon_{n})
\right\}
\nonumber
\\
& &
\qquad\quad
=\sum_{\sigma}\sum_{n}\frac{1}{N_L}\sum_{\mib k}
\left\{
(\varepsilon_{k}-\mu)G^{0}_{\mib k}(i\epsilon_{n})
\frac{{\rm d}}{{\rm d}T}\left[TG_{\mib k}(i\epsilon_{n})\Sigma_{\mib k}(i\epsilon_{n})\right]
\right.
\nonumber
\\
& &
\left.
\qquad\qquad\qquad\qquad\qquad\qquad
+\frac{{\rm d}}{{\rm d}T}\left[(\varepsilon_{k}-\mu)G^{0}_{\mib k}(i\epsilon_{n})\right]\cdot
TG_{\mib k}(i\epsilon_{n})\Sigma_{\mib k}(i\epsilon_{n})
\right\}
\nonumber
\\
& &
\qquad\quad
=\sum_{\sigma}\sum_{n}\frac{1}{N_L}\sum_{\mib k}
\frac{{\rm d}}{{\rm d}T}
\left[T(\varepsilon_{k}-\mu)G^{0}_{\mib k}(i\epsilon_{n})
G_{\mib k}(i\epsilon_{n})\Sigma_{\mib k}(i\epsilon_{n})\right].
\label{eq:S8}
\end{eqnarray}

The thermodynamic potential $\Omega_0$ for the free electrons is given by {Eq.\ (\ref{eq:A3})} 
and is calculated straightforwardly as\cite{Luttinger}:
\begin{eqnarray}
%\Omega_0&=&\sum_\sigma T\sum_{n}\sum_{\mib{k}}e^{i\epsilon_n\delta}
%\ln\left[G^{(0)}_{\mib{k}}(i\epsilon_n)\right]\non\\
%&=&\sum_\sigma\frac{-1}{2\pi}\oint dz f(z) e^{z\delta}\ln\left[G^{(0)}_{\mib{k}}(z)  \right]
\Omega_0=-T\sum_\sigma\frac{1}{N_L}\sum_{\mib{k}}\ln\left[1+e^{-(\epsilon_{\mib{k}}-\mu)/T}\right].
\label{Omega0}
\end{eqnarray}
Therefore, $C_{0}=-T({\rm d}^{2}\Omega_{0}/{\rm d}T^{2})$ is given by 
\begin{eqnarray}
C_0=\sum_\sigma\frac1{N_L}\sum_{\mib{k}}\left(\frac{\epsilon_{\mib{k}}-\mu}{T}\right)^2
\frac{e^{(\epsilon_{\mib{k}}-\mu)/T}}{\left[e^{(\epsilon_{\mib{k}}-\mu)/T}+1\right]^2}.
\label{c0}
\end{eqnarray}

Finally, the total specific heat $C$ is expressed as
\begin{eqnarray}
C=C_0+\tilde{C}{+C^{\prime}_{1}+C^{\prime}_{2}},
\label{c3}
\end{eqnarray}
where ${C^{\prime}_{1}}+{\tilde C}$ is given by Eq. (\ref{eq:S8}).  Namely, 
\begin{equation}
C=C_{0}+\sum_{\sigma}\frac{1}{N_L}\sum_{\mib k}
\frac{{\rm d}}{{\rm d}T}
\left[T\sum_{n}(\varepsilon_{k}-\mu)G^{0}_{\mib k}(i\epsilon_{n})
G_{\mib k}(i\epsilon_{n})\Sigma_{\mib k}(i\epsilon_{n})\right]{+C^{\prime}_{2}}.
\label{eq:S9}
\end{equation}
Because of the relation $G^{0}G\Sigma=G-G^{0}$, the second term of Eq.\ (\ref{eq:S9}) is 
given by 
\begin{equation}
\frac{{\rm d}}{{\rm d}T}\left[
\sum_{\sigma}\frac{1}{N_L}\sum_{\mib k}T\sum_{n}(\varepsilon_{k}-\mu)G_{\mib k}(i\epsilon_{n})
\right]
-\frac{{\rm d}}{{\rm d}T}\left[
\sum_{\sigma}\frac{1}{N_L}\sum_{\mib k}T\sum_{n}(\varepsilon_{k}-\mu)G^{0}_{\mib k}(i\epsilon_{n})
\right].
\label{eq:S10}
\end{equation}
The last term of Eq.\ (\ref{eq:S10}) is easily transformed as  
\begin{eqnarray}
& &
-\frac{{\rm d}}{{\rm d}T}\left[
\sum_{\sigma}\frac{1}{N_L}\sum_{\mib k}T\sum_{n}(\varepsilon_{k}-\mu)G^{0}_{\mib k}(i\epsilon_{n})
\right]
\nonumber
\\
& &
\qquad\quad
=-\frac{{\rm d}}{{\rm d}T}\left[
\sum_{\sigma}\frac{1}{N_L}\sum_{\mib k}(\varepsilon_{k}-\mu)\left(-\frac{1}{2}\right)
\tanh\frac{\varepsilon_{k}-\mu}{2T}\right]
\nonumber
\\
& &
\qquad\quad
=-\sum_{\sigma}\frac{1}{N_L}\sum_{\mib k}\left(\frac{\varepsilon_{k}-\mu}{T}\right)^{2}
\frac{e^{(\epsilon_{\mib{k}}-\mu)/T}}{\left[e^{(\epsilon_{\mib{k}}-\mu)/T}+1\right]^2}
\nonumber
\\
& &
\qquad\quad
=-C_{0},
\label{eq:S11}
\end{eqnarray}
where we have used the expression for $C_{0}$ in Eq.\ (\ref{c0}).  

Therefore, the specific heat [Eq.\ (\ref{c3})] is reduced to 
\begin{equation}
C_{\mu}=\frac{{\rm d}}{{\rm d}T}\left[
\sum_{\sigma}\frac{1}{N_L}\sum_{\mib k}T\sum_{n}(\varepsilon_{k}-\mu)G_{{\mib k}\sigma}(i\epsilon_{n})
\right]{+C^{\prime}_{2}},
\label{eq:S12}
\end{equation}
where we explicitly {represent} the specific heat as $C_{\mu}$ to indicate clearly 
that it is obtained under the condition of a fixed chemical potential $\mu$, 
{and $C^{\prime}_{2}$ is given by Eq.\ (\ref{c12})}.  
Equation (\ref{eq:S12}) is compact and a natural extension of that of $C_{0}$, 
Eq.\ (\ref{eq:S11}) or (\ref{c0}), and is simply the expression in Eq.\ (\ref{eq:result}).  
However, it is much easier to use Eq.\ (\ref{eq:S9}) for numerical 
calculations because the second term Eq.\ (\ref{eq:S9}) more rapidly converges 
than Eq.\ (\ref{eq:S12}) concerning the summation over the Matsubara frequencies $\epsilon_{n}$'s.   

\section{{Concluding Remarks}}
Starting from the Luttinger-Ward expression for the thermodynamic potential, 
we have derived a compact and 
exact formula for the specific heat $C_{\mu}$ (with the fixed chemical potential $\mu$) 
in single-component 
interacting fermion systems in the form of Eq.\ (\ref{eq:S9}) or (\ref{eq:S12}).  An advantage 
of using these expressions is that the specific heat $C_{\mu}(T)$ can be calculated 
at any temperature $T$ 
in strongly correlated fermion systems provided that the single-particle Green function is given by 
some means, e.g., dynamical mean-field theory~\cite{DMFT} and $1/N$-expansion theory of \`a la 
Nagoya,~\cite{Tsuruta}.  

{
Since the above derivation is exact, the specific heat $C_{\mu}(T)$ corresponding to to 
the leading order of $T$ 
in the low-temperature limit must coincide with the expression given by Luttinger~\cite{Luttinger2}. 
Verifying this equivalence is left for a future study.  
}

The expression for entropy is given by integrating the relation 
$(\partial S/\partial T)_{\mu}=C_{\mu}(T)/T$, 
with $C_{\mu}(T)$ given by Eq.\ (\ref{eq:S12}), with respect to the temperature $T$ as follows: 
\begin{eqnarray}
& &
S(T)=\int_{0}^{T}{\rm d}T^{\prime}\frac{1}{T^{\prime}}\frac{{\rm d}}{{\rm d}T^{\prime}}
\left[
\sum_{\sigma}\frac{1}{N_L}\sum_{\mib k}T^{\prime}\sum_{n}
(\varepsilon_{k}-\mu)G_{{\mib k}\sigma}(i\epsilon^{\prime}_{n};T^{\prime})
\right]
\nonumber
\\
& &\qquad\qquad
{+\frac{1}{2}
\int_{0}^{T}{\rm d}T^{\prime}
\sum_{\sigma}\frac{1}{N_L}\sum_{\mib k}
\frac{{\rm d}}{{\rm d}T^{\prime}}\sum_{n}
\left[\Sigma_{{\mib k}\sigma}(i\epsilon^{\prime}_{n};T^{\prime})
G_{{\mib k}\sigma}(i\epsilon^{\prime}_{n};T^{\prime})
\right]},
\label{eq:S15}
\end{eqnarray}
where $\epsilon^{\prime}_{n}\equiv (2n+1)\pi T^{\prime}$ 
{, and explicit $T$ dependences (other than that through the Matsubara 
frequencies) of the Green function and the self-energy are expressed as 
$G_{{\mib k}\sigma}(i\epsilon_{n};T)$ 
and $\Sigma_{{\mib k}\sigma}(i\epsilon_{n};T)$, respectively.}

Although we have developed the formalism for calculating the specific heat on the basis of 
the single-component fermion system with an instantaneous two-body interaction,
the Luttinger-Ward formalism for the thermodynamic potential is known to be generally applied 
to other systems such as the interacting Bose systems and the electron-phonon interacting 
systems.~\cite{Abrikosov}

\section*{Acknowledgment}
This work is supported by a Grant-in-Aid for Scientific 
Research (No.25400369) from Japan Society for the Promotion of Science.

\end{document}